\documentclass[aps,prl,twocolumn,showpacs,superscriptaddress]{revtex4}
\usepackage{bm}
\usepackage{graphicx}
\usepackage{epstopdf}

\newcommand{\beq}{\begin{equation}}
\newcommand{\eeq}{\end{equation}}

\begin{document}
\title{Toward a topological scenario for high-temperature superconductivity
of copper oxides}
        \author{V.~A.~Khodel}
	\affiliation{National Research Centre Kurchatov
		Institute, Moscow, 123182, Russia}
	\affiliation{McDonnell Center for the Space Sciences \&
		Department of Physics, Washington University,
		St.~Louis, MO 63130, USA}
		\author{J.~W.~Clark}
	\affiliation{McDonnell Center for the Space Sciences \&
		Department of Physics, Washington University,
		St.~Louis, MO 63130, USA}
         \affiliation{ Centro de Investiga\c{c}\~{a}o em Matem\'{a}tica
                e Aplica\c{c}\~{o}es, University of Madeira, 9020-105
                Funchal, Madeira, Portugal}
	\author{M.~V.~Zverev}
	\affiliation{National Research Centre Kurchatov
		Institute, Moscow, 123182, Russia}
	\affiliation{Moscow Institute of Physics and Technology,
		Dolgoprudny, Moscow District 141700, Russia}

\begin{abstract}
The structure of the joint phase diagram demonstrating high-$T_c$
superconductivity of copper oxides is studied on the basis of
the theory of interaction-induced flat bands.
Prerequisites of an associated topological rearrangement of the Landau
state are established, and related non-Fermi-liquid (NFL) behavior of the
normal states of cuprates is investigated.  We focus on manifestations
of this behavior in the electrical resistivity $\rho(T)$, especially
the observed {\it gradual crossover} from normal-state $T$-linear
behavior $\rho(T,x)=A_1(x)T$ at doping $x$ below the critical value
$x_c^h$ for termination of superconductivity, to $T$-quadratic
behavior at $x>x_c^h$ \cite{hussey1}, which is incompatible with
predictions of the conventional quantum-critical-point scenario.
It is demonstrated that  at $x<x^h_c$, in agreement with available
  experimental data, the coefficient $A_1( x)$ is decomposed into the
   product of two factors, one of which changes linearly with doping $x$,
    while the second   is universal, being of the Planckian form \cite{zaanen0,tai}.

\end{abstract}


\maketitle

\begin{figure}[t]
\begin{center}
\includegraphics[width=0.8\linewidth] {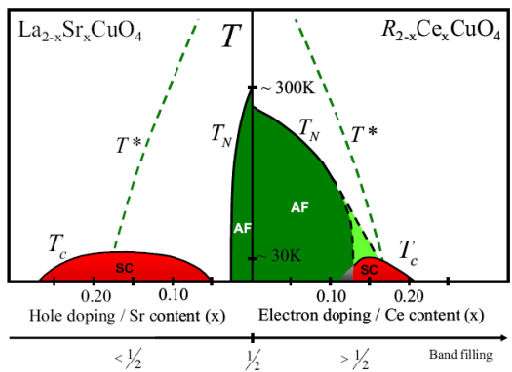}
\end{center}
\vskip -0.3 cm
\caption{Joint phase diagram of copper oxides.  From
Armitage et al. \cite{greene2}. Solid curves labeled $T_c$ and $T_N$
track critical and N\'eel temperatures, respectively; dashed curves
labeled $T^*$ mark approximate extent of pseudogap phases adjacent to
antiferromagnetic (AF) domains.}
\label{fig:joint_pd}
\end{figure}

{\bf Introduction.}
The phenomenon of high-temperature superconductivity (HTSC) was discovered
in two-dimensional electron systems of cuprates in 1986 \cite{bednorz}.
The joint phase diagram \cite{greene2} of hole- and electron-doped
compounds in the temperature-doping ($T$-$x$) plane, reproduced in
Fig.~\ref{fig:joint_pd}, shows two respective superconducting domes,
whose external boundaries are situated at $x^h_c\simeq 0.3$ on the
hole-doped side and $x^e_c\simeq 0.2$ on the electron-doped side.  Their
splitting at $x = 0$ is triggered by an intruding antiferromagnetic
insulating Mott phase.

The origin of HTSC still remains a central issue of condensed matter
physics, unresolved thirty years after its discovery.  Apart from the
puzzling arrangement of the phase diagram and the superconducting phase
{\it per se}, attention has centered on the extraordinary non-Fermi-liquid
(NFL) behavior that has been well documented in the normal states of
cuprates during the last decade.  The dominant attempts to understand
this challenging behavior
(e.g.~\cite{mbp,mu,sca,loch,steglich}),
already promoted in the 1990s, postulate that its source lies in critical
antiferromagnetic fluctuations that generate strong interactions between
quasiparticles near the Fermi surface. In the course of time, this picture
was embodied in a more sophisticated theoretical framework: the quantum
critical point (QCP) scenario, with the driving QCPs usually identified
as $T=0$ end points of lines of {\it second-order} phase transitions, as
a rule of the N\'eel type.

In the QCP scenario, NFL behavior is ascribed to the divergence of the density
of states $N(0)=p_FM^*/\pi^2$ caused by vanishing of the quasiparticle weight
$z=\left(1-(\partial\Sigma(p_F,\varepsilon)/\partial\varepsilon)_0\right)^{-1}$
in single-particle states, stemming from divergent contributions to the
derivative $\left(\partial\Sigma(p_F,\varepsilon)/\partial\varepsilon\right)_0<0$
induced by critical fluctuations \cite{doniach}.  {\it Conclusion}: the Landau
quasiparticle picture ceases to apply, ``Quasiparticles get heavy and die''
\cite{ramaz}.

However, spin fluctuations having the antiferromagnetic vector ${\bf Q}
=(\pi,\pi)$ associated with N\'eel transitions are known to be irrelevant
to the divergence of $M^*$~\cite{chgy}.  Additionally, there exist
generic theoretical objections against vanishing of the $z$ factor at
points of second-order phase transitions \cite{kcz2010}.  These objections,
supported by experimental findings, become especially strong on the
external boundary $x^h_c$ of the superconducting domain
where, with certainty, the quasiparticle picture continues to hold
\cite{yoshida1,yoshida2,yoshida3,bozovic2016,boz1}.
 Furthermore, it has been
stressed in Ref.~\cite{coleman} that quantum criticality develops a $T=0$
phase transition into a state of {\it broken symmetry}.  Given the
observation that on both sides of the joint phase diagram, the
superconducting domes border on {\it conventional Fermi Liquid} (FL)
{\it phases} \cite{greene2}, this fact definitely rules out its relevance
to the part of the phase diagram of cuprates adjacent to the critical
doping values $x^e_c$ and $x^h_c$ considered in this paper.

In contrast, there are different rearrangement options that affect
{\it single-particle} (sp) rather than {\it collective} degrees of
freedom, in which only the topology of the Fermi surface is changed,
without breaking any symmetry inherent in the original ground state.
The crucial consequence of the proposed topological rearrangement of the
ground state lies in the fact that the quasiparticle pattern is preserved
through the transition, implying that the weight $z$ remains {\it finite}.

The first such {\it topological scenario} for the reconstruction of the
Fermi surface beyond a topological critical point (TCP) where the Landau
state loses its stability, was envisioned and studied by
I.\ M.\ Lifshitz in a seminal article published in 1960 \cite{lifshitz}.
His scenario reduces to a change of the number of sheets of the Fermi
surface, with the Landau occupation numbers for quasiparticle states
remaining intact at 0 and 1. Maintenance of the latter property
implies that in Lifshitz transitions, the minimum of the ground-state
energy $E(n)$ is still achieved on the {\it boundaries} of the domain
${\cal D}$ of all possible distribution functions $n({\bf p})$ satisfying
$0 \leq n({\bf p}) \leq 1$, as in standard FL theory \cite{lan}.

However, there also exists a more profound topological rearrangement of the
Landau state, in which the minimum of $E(n)$ with respect to $n$ occurs
at an {\it internal point} of the domain ${\cal D}$.  Although traditional
Landau FL theory itself is no longer applicable for such systems, its basic
ingredient, namely the Landau postulate \cite{lan} that the ground-state
energy $E$ of a Fermi liquid is a functional of the quasiparticle momentum
distribution $n({\bf p})$, still applies. (For a proof, see
Ref.~\cite{jltp2017}). In this situation, a new ground-state quasiparticle
momentum distribution, hereafter denoted $n_*({\bf p})$, is found with
the aid of the variational condition \cite{ks}
\begin{equation}
\frac{\delta E}{\delta n({\bf p})}-\mu=0 , \quad {\bf p}\in \Omega  .
\label{varfc}
\end{equation}
Since the left side of Eq.~(\ref{varfc}) is just the quasiparticle
energy $\epsilon({\bf p})$, this condition now implies the emergence,
in the momentum region $\Omega$, of a dispersionless, compact part
of the spectrum $\epsilon({\bf p})$, originally dubbed the {\it fermion
condensate} (FC), by analogy with the quark condensate, a facet of the
QCD sum rules derived by Shifman, Vainshtein, and Zakharov~\cite {svz}.
Other names commonly associated with the dispersionless portion of the
spectrum $\epsilon({\bf p})$ are {\it flat band} \cite{vol1} and {zero-energy
mode} \cite{wilczek}.

Further comment on the physical rationale for the term fermion condensation is
in order, since this term has lately been introduced in quite different
contexts.  (In particular, the dramatic experimental demonstration of
fermion condensation in cold atomic gases \cite{greiner} refers to
condensation of fermion {\it pairs}, rather than individual fermions.)
As background, let us recall that quantitative description of the
{\it condensation} of a vapor of particles to a quantum liquid is
inherent in the Hohenberg-Kohn theorem \cite{HK}, which declares that
the ground-state energy $E$ is a unique functional of the density
$\rho({\bf r})$, its equilibrium distribution being derived from the
equation
\beq
\frac{\delta E(\rho)}{\delta \rho({\bf r})}=\mu ,
\label{hkc}
\eeq
analogous to Eq.~(\ref{varfc}).
We note further that fermion condensation shares with Bose-Einstein
condensation (BEC) the property that density of states $N(\varepsilon)$
possesses a singular term $\propto \delta(\varepsilon)$, although
phase coherence is present only in the BEC case.  This implies that
in both situations, there is macroscopic occupation of the sp
state of zero energy relative to the chemical potential, the Pauli
principle being preserved in the case of fermion condensation
because the FC particles have different momenta within the FC domain
${\bf p} \in \Omega$.

The prime objection against the original model of fermion condensation
was posed by P.\ Nozi\`eres~\cite{noz}, who addressed, within perturbation
theory, the decay of sp excitations in the presence of a FC (which, {\it
inter alia},
 could  play a key role in many puzzling phenomena of HTSC, e.g.
   the presence of a second, non-BCS gap in the electron spectrum).
In fact, Nozi\`eres made fundamental contributions to the theory of
fermion condensation; in particular, he was the first to develop a
nontrivial finite-temperature model.
  However, his perturbative
method for evaluation of damping effects in systems exhibiting a FC,
which predicted a catastrophic decay rate, is flawed.  A more sophisticated
analysis of the problem
 has provided a resolution of this
issue that supports the FC concept~\cite{jltp2017}.  Consistent with
this conclusion is the recent observation
 of a large and
well-defined Fermi surface in the overdoped LSCO compounds, in both
photoemission and quantum-oscillation experiments~
\cite{damas,vignolle}.

Accordingly, it is relevant to take note of several earlier insights
derived from the FC concept
\cite{ks,vol1,noz,yak2005,prb2008,prb2013,book},
 beginning with the enigmatic behavior of
the low-temperature entropy $S(T\to 0)$ observed in strongly correlated
Fermi systems, ranging from heavy-fermion metals such as CeCoIn$_5$
\cite{steglichce}
to two-dimensional $^3$He films~\cite{saunders}, which are understood
within the FC scenario \cite{annals,shaghe,prb2012}, but as yet
not otherwise.  There also exists compelling experimental evidence
\cite{PRL112} for the validity of the FC scenario from the observation
of its discrete analog: the merging of neighboring sp states in finite
and inhomogeneous Fermi systems, as predicted in Ref.~\cite{merging}.
In addition, there is the upsurge and impending torrent in experimental
and theoretical studies stimulated by the recent discovery of non-BCS
superconductivity in twisted multi-layer graphene, which ``exhibits
ultraflat bands near charge neutrality'' \cite{natur2018}, predicted
more than 25 years ago~\cite{ks}.

The FC scenario aptly describes diverse critical behavior of strongly
correlated electron systems of solids, as documented in
  Refs.~\cite{prb2012,prb2013,prb2016,book,jetpl2015,kcz2017}.
In the present article, we
continue along the same line, focusing attention {\it solely} on that
part of the joint phase diagram of Fig.~1 near its external boundaries
$x^e_c$ and $x^h_c$.  Pseudogap effects associated with reconstruction
of the Fermi surface are nonexistent there, greatly facilitating analysis.
First  we discuss prerequisites for breakdown of the topological
stability of the Landau state. Precursors of its topological rearrangement
beyond the TCP are then identified and examined, in turn, for the
homogeneous electron liquid and for the 2D electron liquid in cuprates.
Finally, we address and resolve the observed dichotomy of the resistivity
of the normal states of high-$T_c$ cuprates, which is inexplicable within
the QCP scenario.

{\bf Generic prerequisites for a topological rearrangement of the
Landau state.}
As is well known \cite{lifshitz,vol1}, violation of the topological
stability of the Landau state is signaled by a change of the number
of roots of the equation
\beq
\epsilon({\bf p},x_c)=0
\label{tope}
\eeq
that determines the structure of the Fermi surface.  One sees
that the topological rearrangement comes into play at a
{\it certain point} ${\bf p}_c$ in momentum space, rather than
ubiquitously as in the QCP scenario.

With restriction to the nominal Fermi surface of the system,
Eq.~(\ref{tope}) may be recast in a more convenient form based on
the FL formula $\epsilon({\bf p}\to {\bf p}_c)=v_F({\bf p}_c)\Delta p $,
where $\Delta p$ is the distance between the momentum ${\bf p}$ and
its critical value ${\bf p}_c$. Inserting this relation into
Eq.~(\ref{tope}), it takes the form
\beq
v_F({\bf p}_c,n_c) =0  .
\label{topv}
\eeq
in terms of the Fermi velocity $v_F$.  As an illustration, we may consider
the 2D homogeneous electron liquid of MOSFETs, with
$p_c=p_F$ and $v_F=p_F/M^*$.  Then, according to Eq.~(\ref{topv}),
the TCP emerges at a critical density $n=n_c$ where $M^*(n)$, given by
the FL relation
\beq
M/M^*(n)= 1-f_1(n)M/2\pi ,
\label{meffr}
\eeq
is divergent.   Then, in the vicinity of the TCP density
  $n_c=7.9\times 10^{10}~{\rm cm}^{-2}$
\cite{mokashi}, the first harmonic
$f_1$ of the Landau interaction function $f$ changes smoothly with $n$
to yield
\beq
M/M^*(n)\propto n-n_c ,
\label{lin}
\eeq
 in agreement with available experimental data \cite{sk}. At the same
time,  if the critical density  $n_c$ were associated with the QCP  scenario,
 the Landau quasiparticle picture  would be destroyed in the immediate vicinity
  of $ n_c$, in stark contrast to the real experimental   situation.

Noteworthy, in microscopic  calculations of  the single-particle spectrum
 of homogeneous 3D electron liquid~\cite{zhetp}, an additional root of Eq.~(\ref{tope}),
 emerges at $p_c(n_c)=0.6p_F$, rather than at the nominal Fermi surface,
 violation of the topological stability  occurring at extremely low density,
 at $r_{sc}\simeq 20$.  Further  reduction of $n_c$ entails the shift of
 the root $p_c$ toward the Fermi momentum $p_F$, and already
  at relatively small variation of $n$,  the TCP momentum $p_c(n)$ attains
   the Fermi surface. In this case, in accord with Eq.(\ref{varfc}), the respective
   rearrangement of the ground state   is accompanied by the   dramatic
    enhancement of the density of single-particle    states,  leading to
      the occurrence of unconventional superconductivity  \cite{ks} where
        the critical    temperature $T_c(n)$  changes linearly
 with  $|n-n_c|$, as in cuprates (see Fig.~1). It is such a behavior
  of $T_c(n)$ that  takes place in doped strontium titanate~\cite{behnia}
  where  the superconducting dome  is located at  $n\leq 3.5 \times 10^{20}$cm$^{-3}$,
   i.e. at $r_s\geq 17$, in accord with   $r_{sc}\simeq 20$,  found above.

{\bf Precursors of the TCP in a homogeneous electron liquid.}
In dealing with transport properties of strongly correlated electron
systems of cuprates, predictions of the two scenarios--QCP and TCP--clash
with each other even on the FL side of the superconducting phase
transition occurring in a homogeneous electron liquid.  To demonstrate
this fact, we examine the resistivity
\beq
\rho(T)=\rho_0+A_1T+A_2T^2 ,
\label{a12}
\eeq
whose nonzero value in metals is ascribed to impurity-induced scattering
and Umklapp processes.  We proceed based on the textbook relation
\beq
\sigma\propto(e^2n\tau)/m^*
\label{trho}
\eeq
for the electric conductivity, in which $m^*$ represents the effective
mass of light carriers and the collision time $\tau$ is expressed in
terms of the conventional Boltzmann integral \cite{pn},
\beq
\tau^{-1}\propto\biggl\{W[n_1n_2(1-n'_1)(1-n'_2)- (1-n_1)(1-n_2)n'_1n'_2]
\biggr\},
\label{coll}
\eeq
where $n(\epsilon)=(1+e^{\epsilon/T})^{-1}$.  The block
$W\propto |\Gamma|^2$ in the collision integral stands for
the transition probability, $\Gamma$ being the scattering amplitude.
The brackets in relation (\ref{coll}) signify integration/summation
over all intermediate momentum and spin variables.

In this section we address the simplest case involving only a single
band crossing the Fermi surface, so that $m^*=M^*$.  Standard but
lengthy algebra \cite{pn} then leads to the behavior
\beq
\tau^{-1}\propto (M^*)^3 T^2   |\Gamma|^2,
\label{coll1}
\eeq
where $|\Gamma|$ is the scattering amplitude averaged over Fermi surface.
Importantly, near the TCP, the scalar part $\Gamma_0$ of the scattering
amplitude $\Gamma$, associated primarily with the zeroth harmonic
$f_0>0$, takes a universal value \cite{kschuck}
\beq
\Gamma_0 \simeq \frac{f_0}{1+f_0N(0)}
   \simeq
M/M^*
\label{gam}
\eeq
due to the TCP divergence of the density of states $N(0)\propto M^*/M$.
The same estimate is obtained for a spin-dependent part of the scattering
amplitude by applying Mermin's sum rule \cite{mermin}.  Upon inserting
these results into Eq.~(\ref{coll1}), we arrive at the relation
$\tau^{-1}\propto M^*$ and thus find
\beq
  \rho(T,n)\propto 1/\sigma(T,n)\propto (M^*(n))^2 T^2 \propto  C^2(T,n),
\label{kw}
\eeq
where $C(T,n)\propto M^*(n)T$ is the specific heat.  This result is in
agreement with the empirical Kadowaki-Woods relation $\rho(T)/C^2(T)
= {\rm const.}$ \cite{kw}.

{\bf Precursors of the TCP in the 2D electron liquid of cuprates.}
Galilean invariance, originally employed by Landau in deriving the equations
of FL theory, does not apply to the electron liquid present in solids.
An adjustment involving gauge invariance, introduced by L.\ P.\ Pitaevskii,
permits the basic FL equation to be recast in the form \cite{pit}:
\beq
{\bf v}({\bf p})={\bf v}_0({\bf p})+\int f({\bf p},{\bf p}_1)
\nabla  n({\bf p}_1) \frac {2d^3{\bf p}_1}{(2\pi)^3},
\label{pit}
\eeq
where the $T=0$ quasiparticle momentum distribution $n({\bf p})$ is given
by the standard FL formula $n({\bf p})=\theta(-\epsilon({\bf p}))$. The
quantity ${\bf v}_0({\bf p})$ playing the role of the bare group
velocity is introduced as ${\bf v}_0({\bf p})=z{\cal T}^{\omega}({\bf p})/M$,
where ${\cal T}^{\omega}({\bf p} )=\lim {\cal T}({\bf p};k{\to}0,\omega{\to}0;
kv_F/\omega{\ll}1) $ \cite{pit}, implying that
in homogeneous matter where $z{\cal T}^{\omega}({\bf p})={\bf p}$ \cite{pit},
 Eq.~(\ref{pit}) coincides with the standard Landau equation
\cite{lan}.  Finally,  a
two-dimensional analog of Eq.~(\ref{pit}) takes the form
\beq
v({\bf p})= v_0({\bf p})-\int_C  f({\bf p},{\bf p}_1)\cos\theta
\frac{2p_1(l_1)dl_1}{(2\pi)^2},
\label{il}
\eeq
with $\cos\theta\propto \left({\bf p}\cdot{\bf v}({\bf p}_1)\right)$.

\begin{figure}[t]
\begin{center}
\includegraphics[width=0.9\linewidth,height=0.45\linewidth]
{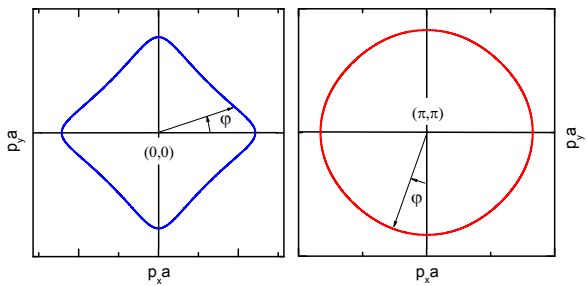}
\end{center}
\vskip -0.3 cm
\caption{Schematic illustration
of the TCP Fermi line. Left panel:
hole-overdoped  LSCO. The Fermi line forms a  square with rounded
corners centered at the point $(0,0)$ \cite{yoshida2}.
 Right panel: electron-overdoped LCCO. In this case, the Fermi line
 is a circle centered at the point $(\pi,\pi)$. }
\label{fig:Fermi_line}
\end{figure}

The integration path $C$, which coincides with the Fermi line itself, is
unknown, except for those few 2D compounds for which detailed angle-resolved
photo-electron spectroscopy (ARPES) data are available.  Among these are
La$_{1-x}$Sr$_x$CuO$_4$ compounds (e.g., see
Refs.~\cite{yoshida1,yoshida2,yoshida3}), in which the Fermi line looks
like a square with rounded corners, as in the left panel of
Fig.~\ref{fig:Fermi_line}.

It is instructive to compare results of calculations based on Eq.~(\ref{il})
in two relevant cases: (i) $ {\bf p}={\bf p}_a$, where the momentum ${\bf p}$
lies on the axes, and (ii) ${\bf p}={\bf p}_d$, where it is located on the
zone diagonals.  For the sake of simplicity, bare Fermi velocities $v_{0d}$
and $v_{0a}$ are evaluated with the aid of tight-binding spectra
$\epsilon^0({\bf p})=\epsilon_0-2t(\cos p_xa+\cos p_y a)
-4t'\cos p_xa\cos p_ya -2t''(\cos 2p_xa+\cos 2p_ya) $,
for which $v_{0a}<v_{0d}$. After some algebra one then finds
\beq
v_{Fd}=v_{0d}-F_d, \quad v_{Fa}=v_{0a}-F_a\sqrt{2} ,
\label{vlsco}
\eeq
where $F_{d,a}>0$ stands for the corresponding integral on the right
side of Eq.~(\ref{il}) containing the repulsive interaction function
$f$.  If the interaction $f$ were weak, the Lifshitz-Volovik condition
(\ref{tope}) would be met at doping $x_B\simeq 0.2$, where the LSCO Fermi
line reaches boundaries of the Brillouin zone \cite{yoshida2}, and hence
the group velocity $v_{0a}(x\to x_B)$ vanishes to provide for logarithmic
 divergence of the density of states
\beq
N(0,x)\propto
\int \frac{d\phi}{v_F(\phi,x)} \propto  \ln |x-x_B| .
\label{dea}
\eeq
Such a  divergence can be treated within a renormalization-group
(RG) formalism \cite{kats1}.  The results demonstrate a flattening of the
sp spectrum $\epsilon({\bf p},x\simeq x_B)$, in accord with behavior
dictated by Eq.~(\ref{varfc}).  However, in the realistic case, one has
$F_a\simeq F_d\simeq 1$.  Then according to Eq.~(\ref{vlsco}), violation
of the topological stability of the Landau state exhibits itself
predominantly near corner points of the Fermi line, rather than at the
zone boundaries, so that both  Eq.~(\ref{vlsco}) and the RG formalism
cease to be applicable.

Altogether, these considerations imply that on the FL side of the TCP, the
total LSCO electron system is separated into two subsystems, consisting of
(i) nodal-region light carriers, specified by the averaged effective mass
$m^*_{av}(x)$, whose properties remain almost unchanged through the
topological transition, and (ii) antinodal-region heavy carriers, identified
by the averaged effective mass $M^*_{av}(x)$, whose value is enhanced
toward $x^h_c$.  The light carriers contribute predominantly to the electric
current, whereas the heavy carriers serve to enhance the inverse collision
time $\tau^{-1}$, whose value becomes:
  \beq
 \tau^{-1}\propto m^*_{av}(M^*_{av})^2 T^2   |\Gamma|^2.
 \label{coll2}
 \eeq

In evaluating the heavy effective mass $M^*_{av}$ associated with
manifestation of NFL behavior in $A_2(x)$, one must recognize that
that the Fermi velocity $v_F(\phi,x^h_c)$ vanishes at the single point
$\phi=0$, while remaining finite at $\phi>0$.  This implies that the heavy
carriers emerge in a narrow momentum region adjacent to the corners of
the Brillouin zone, suppressing their impact accordingly. Furthermore,
as we will see, the enhancement of $M^*_{av}$ depends crucially on the
group velocity itself.   Indeed, according to Eqs.~(\ref{il}) and
(\ref{vlsco}), the behavior of $v(\phi)$ obeys the formulas
\begin{eqnarray}
v_F(\phi,x)&-&v_F(\phi,x_c^h)\propto x-x_c^h ,\nonumber\\
v_F(\phi,x_c^h)&\propto& \phi^2, \quad \phi<\phi_t^h,\nonumber\\
v_F(\phi,x_c^h)&\propto& \phi-\phi_t^h , \quad \phi>\phi_t^h  ,
\label{vfh}
\end{eqnarray}
the last relation being operative at angles $\phi$ greater than a small
transition angle $\phi_t^h$.  Upon inserting Eqs.~(\ref{vfh}) into the
integral (\ref{dea}) and neglecting contributions from the region of
small angles $\phi<\phi_t^h$, we are led to
\beq
N_h(0)=N(0,x\to x^h_c)\propto \ln(|x-x_c^h|+a\phi_t^h) ,
\label{a2h}
\eeq
where $a$ is a numerical constant whose value goes to zero in the
weak-coupling limit $f\to 0$. Otherwise, both $N_h(0)$ and $M^*_{av}$
remain {\it finite}, preventing divergence of the coefficient $A_2(x)$
at $ x^h_c$, {\it contrary} to predictions of the QCP scenario but in
agreement with available experimental data \cite{hussey1}.

\begin{figure}[t]
	\begin{center}
		\includegraphics[width=0.7\linewidth,height=0.6\linewidth]{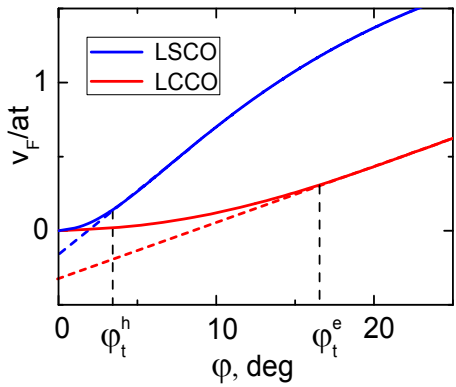}
	\end{center}
	\caption{Fermi velocities $v_F(\phi)$ in units of the product  $at$ at the TCP for
		LSCO (blue solid line) and LCCO (red solid line) compounds. Dashed
		lines indicate the transition angles $\phi_t$ where quadratic
		$\phi$-dependence changes to linear behavior.}
	\label{fig:vF}
\end{figure}

The situation changes on the opposite side of the joint phase diagram,
where the Fermi line tends toward a circular shape (right panel of
Fig.~\ref{fig:Fermi_line}). Indeed, in this case the line integral
in Eq.~(\ref{il}) has the same structure as in a homogeneous 2D electron
liquid; consequently, instead of Eq.~(\ref{vlsco}) one obtains
\beq
v_F(\phi,x)=v_{0F}(\phi)- F_1(x)/2,
\label{vlcco}
\eeq
  where $F_1$ is the corresponding integral in the right side of Eq.~(\ref{il}) proportional to the first harmonic
of the Fourier expansion of the interaction
function $f$ in the 2D Fermi liquid.
The similarity between Eqs.~(\ref{vlsco})
and (\ref{vlcco}) implies that the behavior of the Fermi velocity $v_F(\phi)$
at small $\phi$ and $ x-x_c^e$, given by Eqs.~(\ref{vfh}), remains intact
in going from the LSCO family to the LCCO family. However, the transition
angle $\phi_t$ now becomes large enough (see Fig.~\ref{fig:vF}), so that the
overwhelming contributions to density of states come from the region
$\phi<\phi_t$. This makes the pivotal difference, since upon inserting
Eq.~(\ref{vfh}) into Eq.~(\ref{dea}),
we are led to
\beq
N_e(0,x)\propto M^*_{av}(x) \propto |x-x_c^e|^{-1/2},
\label{mlcco}
\eeq
which differs from the QCP result, where $M^*_{QCP}(x) \propto |x-x_c^e|^{-1}$.
Nevertheless, upon inserting Eq.~(\ref{mlcco}) into Eq.~(\ref{coll2})
one finds
\beq
A_2(x\to x^e_c)\propto (x-x_c^e)^{-1}.
\label{a2e}
\eeq
This behavior, in agreement with experiment, coincides with the
QCP scenario, where $A_2(x)\propto M^*_{QCP}(x)\propto (x-x_c^e)^{-1} $~\cite{greene1},
 at variance with the TCP behavior (\ref{mlcco}).
 
\begin{figure}[t]
\begin{center}
\includegraphics[width=0.9\linewidth,height=0.57\linewidth]{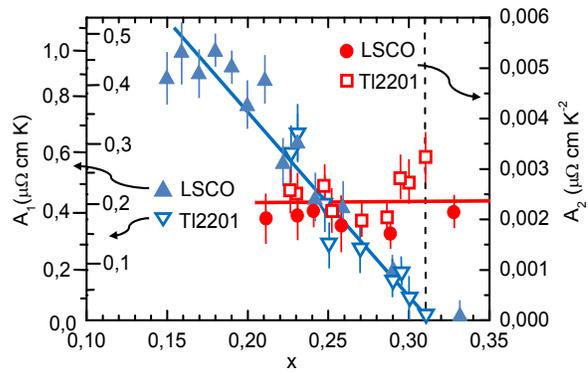}
\end{center}
\vskip -0.5 cm
\caption{Doping dependence of linear-temperature ($A_1$) and quadratic
($A_2$) terms in the resistivity $\rho(T)$ of compounds LSCO and Tl2201.
Left axis: triangles indicate experimental data \cite{hussey1}
for LSCO (scale outside) and Tl2201 (scale inside), while the straight
line shows predicted $A_1\propto |x-x_c|$ with
$x_c^h\simeq 0.3$ and a coefficient chosen
to fit the average trend.  Right axis: circles show experimental
data~\cite{hussey1}, while the horizontal straight line represents
the prediction $A_2\simeq {\rm const.}$, with its value
taken to match experiment.}
\label{fig:a1_a2_LT}
\end{figure}

\begin{figure}[t]
	\begin{center}
		\includegraphics[width=0.9\linewidth,height=0.65\linewidth] {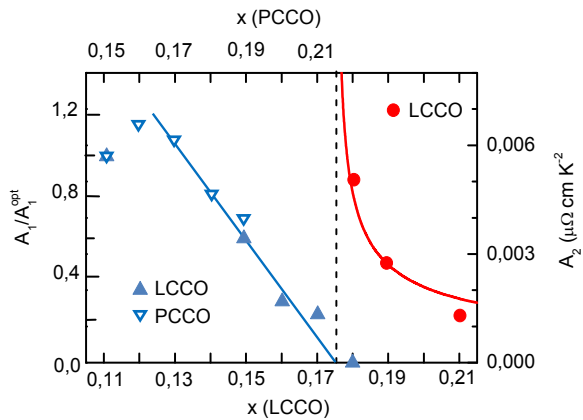}
	\end{center}
	\vskip -0.5 cm
	\caption{Doping dependence of $A_1$, normalized to that of optimal doping,
		for LCCO and PCCO (left side of plot) and likewise $A_2$ for LCCO (right side).
		Left ($x<x_c^e$; $x_c^e\simeq 0.175$ for LCCO and 0.215 for PCCO):
		triangles indicate experimental data \cite{greene1}, while the straight
		line shows TCP prediction $A_1\propto |x{-}x_c^e|$ with a coefficient
		chosen to fit experimental value of reduced $A_1$ for LCCO at $x{=}0.15$.
		Right ($x{>}x_c^e$): circles show experimental data \cite{greene1}, while
		overlaid curve represents TCP result $A_2\propto (x{-}x_c^e)^{-1}$
		with a prefactor taken to reproduce experimental value of $A_2$ for
		LCCO at $x{=}0.19$.
	}
	\vskip -0.3 cm
	\label{fig:a1_a2_LP}
\end{figure}

{\bf Dichotomy in the resistivity of normal states of high-$T_c$
cuprates.}  We are now ready to analyze the NFL behavior of the resistivity
$\rho(T,x)$ of normal states of cuprates, several experimental findings
having attracted broad attention.  First, the dominant feature that must be
acknowledged is that $\rho(T)$ exhibits a predominantly linear dependence on
temperature (note especially Refs.~\cite{hussey1,greene1,greene2,taillefer1}).
Second, instead of collapsing to a single critical point, as the conventional
QCP scenario prescribes, the coefficient $A_1(x)$ is found to grow linearly
in the difference $|x-x_c|$, with a prefactor that is rendolent in its
implications--being {\it universal over different cuprates} with a magnitude
comparable to the Planckian limit~\cite{tai}.  While the linearity of
$\rho(T)$ might be accounted for within the framework of the
Hertz-Millis-Moriya approach \cite{moriya}, its other features, as
explicated above, {\it defy explanation within the fluctuation scenarios}.

Conversely, all these anomalies may be properly clarified within the
framework of the FC concept \cite{ks,vol1,noz,prb2008,jltp2017}
(see Figs.~3 and 4).  Because the pseudogap phase is nonexistent close
to the boundaries in the joint phase diagram~\cite{greene1,yoshida2}, one
can apply the original model of fermion condensation, in which the
FC effective is mass given by $M^*_{FC}(T,x)\propto \eta(x)/T$, with
$\eta(x)\ll 1$ measuring the ratio of the volume in momentum space
occupied by the FC to the total Fermi volume (for details, see
\cite{jetpl2015}).  Upon substituting this relation into Eqs.~(\ref{trho})
and (\ref{coll2}) determining the resistivity $\rho(T)$, one finds
  $\rho(T,x)\propto (ne^2)^{-1} (m^*_{av}(x))^3 M^*_{FC}(T,x)|\Gamma(x)|^2 T^2$,
arriving finally at
\beq
A_1(x)\propto  (ne^2)^{-1} (m^*_{av}(x))^3 |\Gamma(x)|^2  \eta(x) .
\label{tai}
 \eeq
Corrections to this result, being $T$-independent, are proportional to
$\eta^2$.

Bearing in mind that the averaged effective mass $m^*_{av}(x)$ extracted
from the specific heat data of LSCO compounds is large, i.e.,
around $10m_e$ \cite{tai}, one can employ relation ~(\ref{gam}) to
eliminate the product $ (m^*_{av}(x))^2 |\Gamma(x)|^2 $ from Eq.~(\ref{tai}).
The result
\beq
A_1(x)\propto  (ne^2)^{-1} m^*_{av} \eta(x)
\label{tai1}
\eeq
can be conveniently rewritten in
the form
\beq
A_1^{2D}(x)=A_P(x)\eta(x) ,
\label{a12d}
\eeq
 with the so-called Planckian factor~\cite{zaanen0,tai}
\beq
A_P(x)=\alpha \frac{h}{2e^2 T_F(x)}\eta(x) ,
\label{plan}
\eeq
 where $h$ is the Planck constant,
$T_F(x)=p^2_F/2m^*_{av}(x)$, and $\alpha$ is a constant of order
unity. Derived within the framework of the FC concept, Eq.~(\ref{a12d}),
with the {\it doping-dependent factor} $\eta(x)$, is
corroborated by experiment on both sides of the joint phase diagram
 of copper oxides~\cite{tai}. These facts  refute any  connection between  $A_1(x)$,
 and the Planckian limit  of the scattering rate \cite{zaanen0,tai},
giving support to the FC scenario of the HTSC suggested in this article.

We now turn to elucidation of the enigmatic proportionality between
$A_1(x)$ and the critical temperature $T_c(x)$, established in numerous
experimental studies (notably Refs.~\cite{hussey1,greene1,taillefer1}).
The theoretical curve for $T_c(x)$ is found with the aid of a Thouless
criterion, which involves the residues
of the pole of the Cooper vertex part:
\beq
\!{\cal T} ({\bf p},T_c)=\!-\!\!\int\! {\cal V}({\bf p},{\bf p}_1)\frac{\tanh
\frac{\epsilon({\bf p}_1,T_c)}{2T_c}}{2\epsilon({\bf p}_1,T_c)}{\cal T}
({\bf p}_1, T_c) \frac{d^2{\bf p}_1}{(2\pi)^2} .\!\!\!\!
\label{thouless}
\eeq
Overwhelming contributions to the right side of this equation are known
to come from four different FC spots, each of which is associated with its
own saddle point.
Accordingly, Eq.~(\ref{thouless}) is converted to a set of four coupled
equations for the residues ${\cal T}_i\equiv{\cal T}({\bf p}_i)$,
each point ${\bf p}_i$ being coincident with one of the four saddle
points.  This set has nontrivial solutions that do not exist in homogeneous
systems.  Let us focus on the solution ${\cal T}_i=(-1)^i {\cal T}$
having $D$-wave  structure. Bypassing the details of calculation, which
may be found in Refs.~\cite{kcz2017,jltp2017,jetpl2015}, we present the
final result:
\beq
T_c(x)={\cal V}_{f}\eta(x) ,
\label{thou1}
\eeq
where the coupling constant $ {\cal V}_{f}>0$ turns out to be proportional
to the matrix element
${\cal V}_{i,i+1}\equiv {\cal V}({\bf p}_i,{\bf p}_{i+1})$
of the {\it repulsive} effective e-e interaction.  Its unexpected role,
i.e., promoting  superconductivity, is exclusively due to the $D$-wave
gap structure.

Thus, as seen from comparison of Eqs.~(\ref{thou1}) and (\ref{tai1}),
the ratio $A_1(x)/T_c(x)$ turns out to  be doping independent, in
agreement with experiment on both electron-doped and
hole-doped sides of the joint phase diagram of cuprates
\cite{hussey1,greene1,taillefer1}. But importantly: despite
common opinion otherwise, this independence demonstrates that the
dominant contributions to the quantities of most interest stem from
the momentum domain {\it occupied by the FC}, rather than from the
similarity of relevant interactions between quasiparticles.

{\bf Discussion.}
First of all, we observe that the principal difference between the
quantum critical point (QCP) and topological critical point (TCP) scenarios
for HTSC materials is expressed decisively just near the boundaries
of superconducting domes, where $T_c(x)$ terminates by virtue of the
condition $\lambda_D(x_c)=0$, with $x_c$ either $x_c^h$ or $x^e_c$.
Indeed, the QCP-BCS behavior $T_c(x\to x_c)\propto e^{-2/\lambda_D(x)}$
implies that the derivative
\beq
dT_c^{QCP}(x\to x_c)/dx
\propto e^{-2/\lambda_D(x)}
\eeq
of $T_c(x)$ {\it vanishes} at $x=x_c$, whereas experimentally it remains
{\it finite} or even diverges \cite{greene1,greene2,bozovic2016}.  This discrepancy once again
rules out the QCP scenario as an explanation of HTSC of cuprates.

In contrast, the TCP behavior $dT_c(x)/dx\propto d\eta/dx$ associated with
the doping evolution of the FC parameter $\eta(x)$ is free from this flaw.
In several solvable models of fermion condensation as well as numerical
calculations, the parameter $\eta(x)$ grows linearly with $|x{-}x_c|$ or
even more rapidly \cite{prb2008,kcz2010,jltp2017}, in unison with
available experimental data.

The scenario of high-$T_c$ superconductivity introduced herein is
distinguished by the occurrence of {\it several} FC spots, domains in
momentum space where the $T=0$ quasiparticle distribution $n_*({\bf p})$
departs from its FL form, triggering NFL  behavior of strongly
correlated Fermi systems, well documented in diverse experimental studies
of the last decade. It is the
  C$_4$
symmetry of the angular arrangement of
this  departure, which is stable with respect to elevation of $T$ up
to $T_c$, that explains the nontrivial $D$-wave structure of HTSC in
cuprates (for details, see Refs.~\cite{jltp2017,kcz2017}).  In view of the
results obtained in the strong-coupling calculations of Ref.~\cite{kats2}
on a triangular lattice, and especially in light of recent experimental
studies of multi-layer graphene \cite{natur2018}, the correlation between
HTSC and FC spots is presumably quite universal.

Another important result of our analysis bears on explanation of the
so-called anomalous criticality uncovered in conductivity studies
of overdoped LSCO compounds \cite{hussey1}, including the {\it absence}
of a divergence of the coefficient $A_2(x)$ on the FL side of the phase
diagram.  This behavior negates a dictum of the QCP scenario, which
also fails to explain the emergence of a linear-in-$T$ NFL term in the
resistivity $\rho(T)$ at $x<x^h_c$, whose magnitude $A_1(x)$ grows
linearly with the difference $x_c^h-x$~\cite{hussey1}.  In view of
what the existing observations have revealed, it is worth repeating these words
of Cooper et al. \cite{hussey1} nearly a decade ago: ``The strange-metal
physics of hole-overdoped cuprates is associated not with the
presence of a quantum critical point, but instead with a novel
extended phase''--which we have identified as the flat-band state.

In summary, we have demonstrated that in copper oxides, the total domain
of high-temperature superconductivity is confined between two critical
values of the doping level, $x^e_c$ and $x^h_c$, associated with the
boundaries of the momentum region occupied by a fermion condensate, which
emerges due to breakdown of topological stability of the Landau state.
Arguments have been presented, both experimental and theoretical, that
refute the view that extraordinary properties of the cuprates are driven
by quantum critical points.  Our analysis and its results suggest that
while doping of CuO$_2$ planes by electrons or holes may convert a
Mott insulator into a conductor, this observation does not imply that
antiferromagnetic fluctuations promote superconductivity of copper
oxides, but indeed rather the contrary--quite at variance with widespread
belief.

{\bf Acknowledgments.}
We thank V.\ Dolgopolov, L.\ Pitaevskii, R.\ Greene, Ya.\ Kopelevich,
V.\ Shaginyan, and G.\ Volovik for valuable discussions. VAK and JWC
acknowledge financial support from the McDonnell Center for the Space
Sciences.  JWC expresses his gratitude to the University of Madeira and
its branch of Centro de Investiga\c{c}\~{a}o em Matem\'{a}tica e
Aplica\c{c}\~{o}es (CIMA) for gracious hospitality during periods
of extended residence.

\end{document}